\documentclass[showpacs,twocolumn,aps,prb,floatfix,superscriptaddress,noshowpacs]{revtex4}
\usepackage{amsthm} %added by me
\usepackage[mathlines]{lineno}% Enable numbering of text and display math

\usepackage{amsmath,amssymb,graphicx,bm,color,mathrsfs,verbatim,epstopdf,dcolumn,dsfont,hyperref}
\newcommand{\be}{\begin{equation}}
\newcommand{\ee}{\end{equation}}

\begin{document}

\preprint{APS/123-QED}

\title{Stroboscopic versus non-stroboscopic dynamics in the Floquet realization of the Harper-Hofstadter Hamiltonian}%

\author{Marin Bukov and Anatoli Polkovnikov}
\affiliation{Department of Physics, Boston University, 590 Commonwealth Ave., Boston, MA 02215, USA}
\email{mbukov@bu.edu}

\date{\today}

\begin{abstract} 
We study the stroboscopic and non-stroboscopic dynamics in the Floquet realization of the Harper-Hofstadter Hamiltonian. We show that the former produces the evolution expected in the high-frequency limit only for observables which commute with the operator to which the driving protocol couples. On the contrary, non-stroboscopic dynamics is capable of capturing the evolution governed by the Floquet Hamiltonian of any observable associated with the effective high-frequency model. We provide exact numerical simulations for the dynamics of the number operator following a quantum cyclotron orbit on a $2\times 2$ plaquette, as well as the chiral current operator flowing along the legs of a $2\times 20$ ladder. The exact evolution is compared with its stroboscopic and non-stroboscopic counterparts, including finite-frequency corrections.      
\end{abstract}

\maketitle

\section{\label{sec:intro}Introduction}

Proposals using periodic external fields~\cite{breuer_91} to engineer specific properties of matter are currently experiencing an unprecedented flurry of interest. Theoretical models based on Floquet's theorem are being developed to simulate systems in regimes otherwise inaccessible in conventional condensed matter materials~\cite{jaksch_03,mueller_04,eckardt_10,dalibard_11,kitagawa_11,creffield_11,ploetz_11,ploetz_11_interferometry,kolovsky_11,struck_11,struck_12,hauke_12,struck_13,greschner_13,parra-murillo_13,goldman_14,bukov_14}. Experimentally, cold atoms' unique controllability was employed to observe dynamical localisation and phase-coherence in strongly shaken bosonic systems~\cite{lignier_07,sias_07,eckardt_09,zenesini_09,creffield_10,arimondo_12}. This paved the way towards generating extremely strong artificial magnetic fields~\cite{aidelsburger_11} in lattice models, which recently culminated in the realisation of the Harper-Hofstadter model~\cite{aidelsburger_13,miyake_13}, the Quantum Spin Hall Effect~\cite{beeler_13,kennedy_13}, and Floquet topological insulators~\cite{jotzu_14,aidelsburger_14}.    

\begin{figure}[!ht]
\begin{minipage}{\columnwidth}
\includegraphics[scale=0.35]{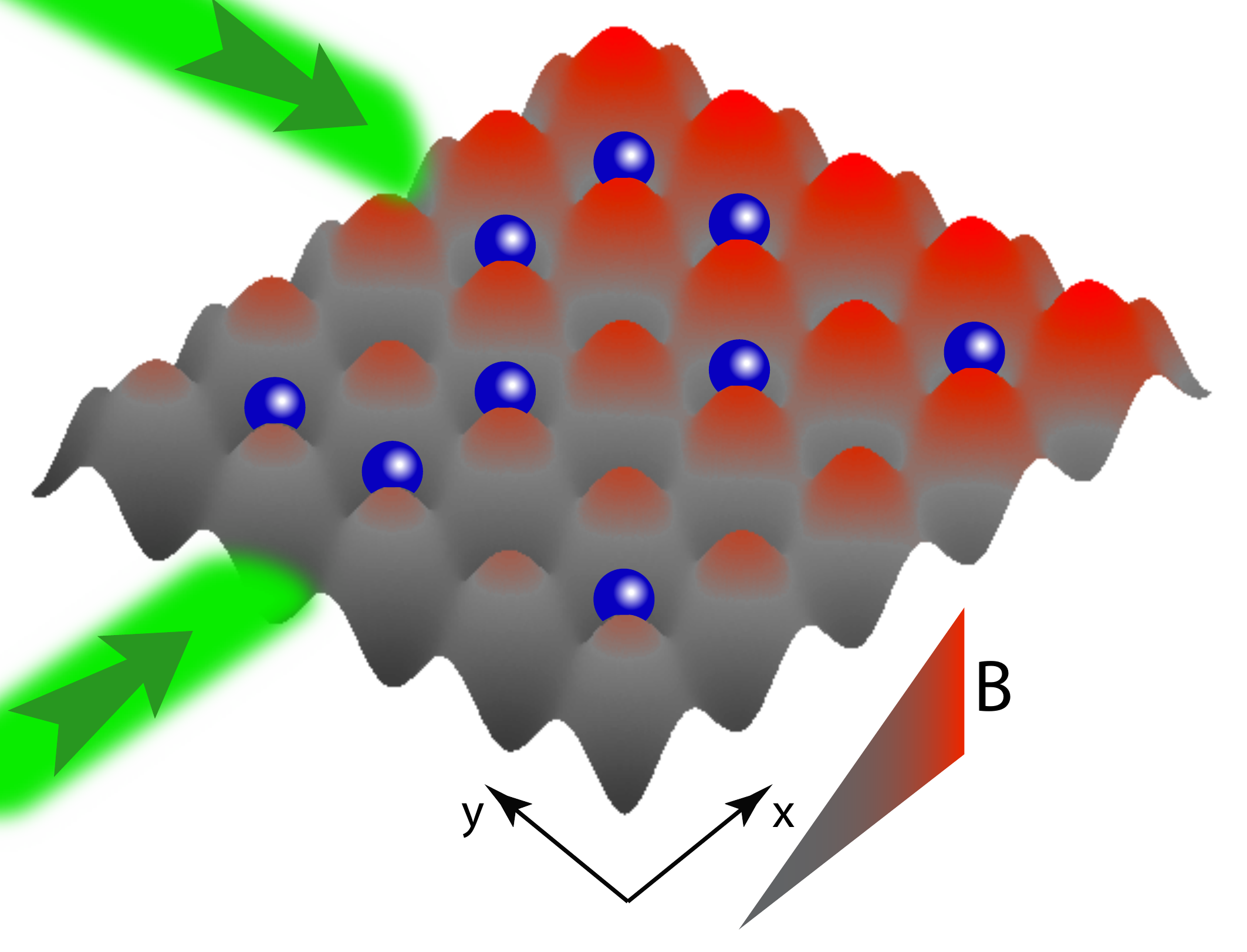}
\end{minipage}
\begin{minipage}{\columnwidth}
\includegraphics[scale=0.35]{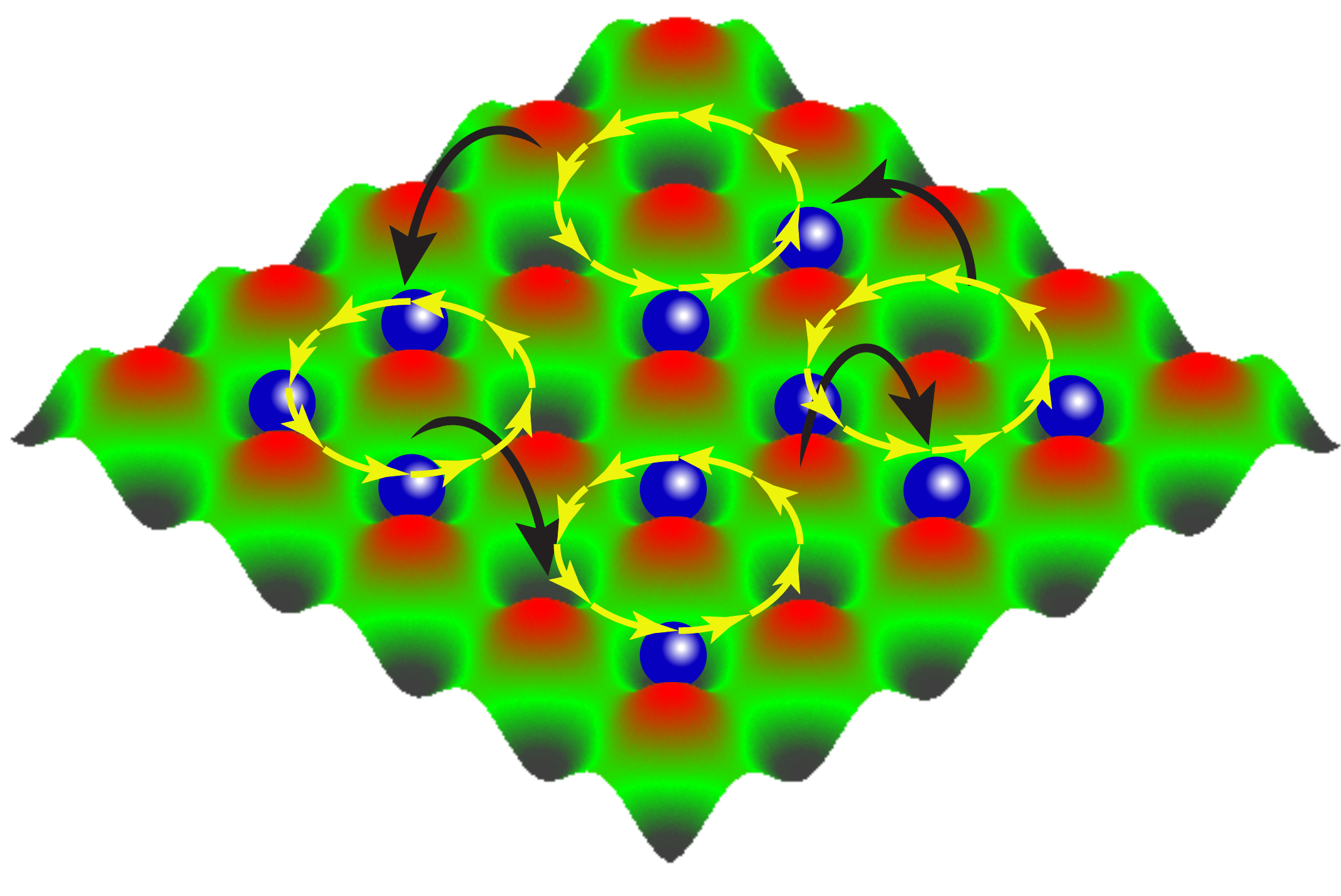}
\end{minipage}
\caption{\label{fig:HH_model}(Color online) The Floquet realisation of the Harper-Hofstadter (HH) model~\cite{harper_55,hofstadter_76}. A magnetic field gradient is applied to inhibit the hopping along the $x$-direction. At the same time, two resonant Raman-Bragg lasers result in a periodic driving protocol with site-dependent phase which couples to the atomic number operator. As a result, in the infinite-frequency limit, the Floquet Hamiltonian coincides with the HH model of lattice bosons in a magnetic field.}
\end{figure}

The success of these experiments triggers a wave of intense study from the theoretical side. In this paper, we carefully analyse the dynamical (Floquet) realisation of the Harper-Hofstadter model~\cite{harper_55,hofstadter_76} of free neutral lattice bosons in a strong artificial magnetic field~\cite{jaksch_03,aidelsburger_13,miyake_13}, c.f.~Fig.~\ref{fig:HH_model}. Since periodically driven systems do not obey the energy conservation law, and given that the equations of motion are not exactly solvable either, this problem poses a considerable challenge already for non-interacting systems. Understanding the behaviour of such systems is a crucial prerequisite for the analysis of strongly driven interacting systems, believed to hold the key to `non-equilibrium thermodynamics'~\cite{arimondo_12} in the high-frequency regime. The purpose of this work is to study in full detail the out-of-equilibrium dynamics of this strongly coupled Floquet system using analytical and numerical tools.

The theoretical analysis of periodically driven systems in the high-frequency regime relies on Floquet's theorem. In essence, it states that the evolution operator of any periodic Hamiltonian $H(t) = H(t+T)$ can be decomposed as 
\begin{equation}
U(t,0) = P(t)\exp(-i H_F t),
\label{eq:floquet_thm}
\end{equation}
where $P(t+T) = P(t)$ is the unitary, periodic Kick operator~\cite{goldman_14}, and $H_F$ is the Floquet Hamiltonian. In the high-frequency limit, $H_F$ governs the slow, and $P(t)$ - the fast evolution. In the original realisation proposed in Ref.~\onlinecite{jaksch_03}, the infinite-frequency Floquet Hamiltonian $H_F$ coincides with the Harper-Hofstadter (HH) model. Often times, theoretical works consider stroboscopic evolution only, which is defined at integer multiples of the driving period $T$. In this paper, we show that this is not enough to measure certain properties of $H_F$, and hence the kick operator $P(t)$ needs to be taken into account. We remark that similar conclusions about the importance of $P(t)$ have recently been drawn in Ref.~\onlinecite{goldman_14}.

By considering a two-dimensional (2D) plaquette and ladder geometries, we compare the exact dynamics of observables, such as the local number and current operators, to that expected from the HH model. Our main objects of interest are the quantum cyclotron orbits and the chiral currents measured in the experiments of Refs.~\onlinecite{aidelsburger_13} and~\onlinecite{atala_14}. We analyse the discrepancies and similarities between the Floquet stroboscopic (FS) evolution (c.f.~Sec.~\ref{sec:FS}) and the Floquet non-stroboscopic (FNS) evolution (c.f.~Sec.~\ref{sec:FNS}) first introduced in Ref.~\onlinecite{bukov_14}. To incorporate finite-frequency effects, we take into account $\Omega^{-1}$-corrections to the Hamiltonian. In particular, we find that, while FS evolution suffices to measure the local number operator, if one wants to measure the chiral current, one has to follow the FNS evolution protocol. We show numerical results for the exact dynamics of these quantities, and make a comparison between the exact, the FS, the FNS and the infinite-frequency curves. We find that the FS evolution reveals the physics of the Floquet Hamiltonian, $H_F$, only for observables which commute with the operator to which the driving couples: in this case - the local number operator. On the other hand, stroboscopic measurements fail to reveal the properties of $H_F$ for non-gauge-invariant quantities, such as the chiral current.  

A necessary condition for cold atom experiments to work is that the driving frequency be smaller than the band gap between the lowest two Bloch bands, or otherwise population of higher bands will occur. This fact is in strong contrast with the infinite-frequency limit assumed in the derivation of the effective Harper-Hofstadter model. Recently, it has been shown that, when taken into account, finite-frequency corrections can generate additional terms, such as next-nearest-neighbour hopping (nnn), diagonal hopping, and a site-dependent chemical potential~\cite{bukov_14}. A similar work classifying the relevant corrections on a one-dimensional chain has been done~\cite{itin_14}, and an equivalent alternative method using the flow-equation approach has also been developed~\cite{verdeny_13}. Hence, any realistic analysis should explore the effects induced by the leading corrections. 

This paper is organised as follows. In Section~\ref{sec:model_corrections} we revisit the derivation of the Harper-Hofstadter model as the infinite-frequency Floquet Hamiltonian of a periodically driven system, and set up the stage for the plaquette and ladder geometries. Sections~\ref{sec:FS} and~\ref{sec:FNS} focus on the physics of the FS and FNS evolution, respectively. In both cases, we show numerical results for the exact evolution of a particle following a quantum cyclotron orbit on a singe plaquette, as well as the Harper-Hofstadter chiral currents flowing along the edges of a $2\times 20$ ladder. Finally, we conclude the analysis in Section~\ref{sec:conclusion}.

\section{\label{sec:model_corrections}Floquet Realisation of the Harper-Hofstadter Hamiltonian}

Consider a system of neutral bosons loaded in a 2D optical lattice. A magnetic field gradient, used to tilt the lattice along the $x$-direction, inhibits the hopping along the $x$-axis. The latter is then restored in a controllable fashion using resonant Raman lasers (running lattice). The site-dependent phase lag of the Raman lasers allows to imprint an arbitrary fixed Peierls phase to the hopping along the $x$-direction (c.f.~Fig.~\ref{fig:HH_model}), breaking time-reversal symmetry. As a result, in the infinite-frequency limit, one obtains an effective magnetic flux per plaquette $\Phi_\square$, the strength of which can be controlled by the running Raman lasers. A detailed description of the experimental set-up can be found in Refs.~\onlinecite{aidelsburger_13,miyake_13}.   

The Hamiltonian of the system reads as

\begin{align}
H^\text{lab}(t) =& H_\text{kin} + H_\text{drive}(t) + H_\text{int},\nonumber\\
H_\text{kin} =& -\sum_{m,n} \left( J_x a^\dagger_{m+1,n}a_{mn} + 
J_y a^\dagger_{m,n+1}a_{mn} + \text{h.c.}\right),\nonumber \\
H_\text{drive}(t) =& \sum_{m,n}\left[\frac{V_0}{2}\sin\left(\Omega t - \phi_{mn} + \frac{\Phi_\square}{2}\right) + \Omega m\right] n_{mn},\nonumber\\
H_\text{int} =& \frac{U}{2}\sum_{m,n} n_{mn}(n_{mn}-1).
\label{eq:H(t)_lab}
\end{align} 
The bare hopping matrix element along the $x$- ($y$-) direction is denoted by $J_x$ ($J_y$), while $U$ is the on-site interaction strength. We denote by $a^\dagger_{mn}$ ($a_{mn}$) the boson creation (annihilation) operator on site $(m,n)$ in the lab frame. Notice that the Raman lasers couple to the local number operator of the system, breaking translational invariance through their phase dependence, $\phi_{mn} = \Phi_\square(n+m)$. We denote the strength of the running lattice by $V_0$, and the driving frequency - by $\Omega$. Pay attention how the magnetic field gradient is locked to the driving frequency, which restores the hopping along the $x$-direction. 

In the following, we assume that the driving frequency $\Omega$ is the largest energy scale in the problem. In realistic experiments, a natural upper bound on this quantity is imposed by the band gap between the lowest two Bloch bands~\cite{aidelsburger_13,miyake_13}. Hence, it is useful to consider corrections to the infinite-frequency Floquet Hamiltonian~\cite{bukov_14}.

In the high-frequency limit, the running lasers oscillate wildly, and one would naively expect that the system feels the time-averaged Hamiltonian. However, the magnetic field gradient is locked to the driving frequency. Therefore, it is not easy to take the limit $\Omega\to\infty$, and find the Floquet Hamiltonian directly in the lab frame. The way out is to perform a time-dependent transformation $V(t)$ \emph{into} a rotating frame~\cite{miyake_13} which amounts to a re-summation of an infinite Magnus sub-series~\cite{magnus_54,bukov_14}: 

\begin{eqnarray}
V(t) &=& \exp\left(i\int^t\mathrm{d}t' H_\text{drive}(t')\right)\nonumber\\
     &=& \text{e}^{ i\sum_{mn}\left[-\frac{V_0}{\Omega}\cos\left(\Omega t - \phi_{mn} + \frac{\Phi_\square}{2} \right) + \Omega mt \right]n_{mn} }.
\label{eq:V_rot_frame}
\end{eqnarray}
This transformation leaves the number operator $n_{mn} = a^\dagger_{mn}a_{mn}$ intact. Although the Floquet Hamiltonian in the rotating frame is unitarily equivalent to the one in the lab frame, the two operators can be different, unless $V(t)$ is stroboscopic~\cite{bukov_14}. This ambiguity is the same as the one associated with the relative phase of the driving protocol (Floquet gauge~\cite{bukov_14}). The relation between the two is given by $H_F^\text{rot} = V(0)H_F^\text{lab}V^\dagger(0)$.

The Hamiltonian in the rotating frame is obtained as

\begin{align}
H^\text{rot}(t) =& -\sum_{m,n}J_x\left(e^{-i\zeta\sin(\Omega t + \phi_{nm}) +i\Omega 
t }a^\dagger_{m+1,n}a_{mn} + \text{h.c.}\right)\nonumber\\
& -\sum_{m,n} J_y\left(e^{-i\zeta\sin(\Omega t + \phi_{nm}) 
}a^\dagger_{m,n+1}a_{mn} + \text{h.c.}\right)\nonumber\\
& +\sum_{m,n} \frac{U}{2} n_{mn}(n_{mn}-1),
\end{align}
where $\zeta = V_0/\Omega\sin(\Phi_\square/2)$ is the dimensionless coupling strength of the driving. 

A rigorously posed high-frequency limit can now be defined in the rotating frame by taking $\Omega\to\infty$ with $\zeta={\rm const}$. The second condition means that the intensity of the running lasers must be of the same order of magnitude as the driving frequency. Was it not for this second condition, the slope of the tilt $\Omega$ would effectively forbid any tunnelling along the $x$-direction, and hopping would remain inhibited even for $\Omega\to\infty$. The infinite-frequency Floquet Hamiltonian \emph{in the rotating frame} coincides with the Harper-Hofstadter model~\cite{harper_55,hofstadter_76}, and is given by

\begin{eqnarray}
H_F^{\text{rot},(0)} &=& -K(\zeta)\sum_{m,n} \left(e^{-i\phi_{mn}}a^\dagger_{m+1,n}a_{mn} + 
\text{h.c.}\right)\nonumber\\
&& -J(\zeta)\sum_{m,n}\left(a^\dagger_{m,n+1}a_{mn} + \text{h.c.}\right) \nonumber\\
&& +\ \ \ \frac{U}{2}\sum_{m,n} n_{mn}(n_{mn}-1),
\label{eq:harper_Heff}
\end{eqnarray}
where the renormalized hopping matrix elements are $K(\zeta) = J_x\mathcal{J}_1(\zeta)$, $J(\zeta) = 
J_y\mathcal{J}_0(\zeta)$, and $\mathcal{J}_\nu$ is the $\nu^\text{th}$ Bessel function. The system, therefore, behaves as if it is subject to a net magnetic field, although the bosonic atoms are in fact not charged~\cite{jaksch_03}. The dimensionless interaction strength $\zeta$ is now a freely adjustable knob, with the help of which one can control the effective model parameters.  

The resulting finite-frequency corrections to the kinetic energy can be understood intuitively in a perturbative fashion. If, to zeroth order (i.e.~to $\mathcal{O}\left(\Omega^{0}\right)$), a boson is allowed to hop between any two nearest-neighbouring (nn) sites, then one can think of the first-order correction, i.e.~to $\mathcal{O}\left(\Omega^{-1}\right)$, as a virtual two-step hopping process. On a two-dimensional lattice, this means that we expect terms representing next-nearest-neighbour (nnn) hopping along both the $x$- and $y$-directions, as well as diagonal hopping. Furthermore, one can imagine a two-hopping process in which a particle hops away and then back to the same site. This will lead to a site-dependent potential term, as it measures the difference of the densities of the two sites involved. The presence of interactions results in the so-called interaction-dependent (or correlation-dependent) hopping $\sim a^\dagger_{m+1,n}a_{mn}(n_{mn} - n_{m+1,n})$. The latter result from a two-step virtual processes where a particle interacts and then hops to the nn site, or vice versa. These terms are diagonal neither in real, nor in momentum space and, therefore, pose a significant challenge to study theoretically. Moreover, their effect is highly dependent on the filling factor of the lattice, as well as the state the system is in. However, these terms are generic for all interacting Floquet models. A detailed discussion of the leading correction to the infinite-frequency Floquet Hamiltonian is given in Ref.~\onlinecite{bukov_14}.

\begin{figure}[b]
\includegraphics[width=\columnwidth]{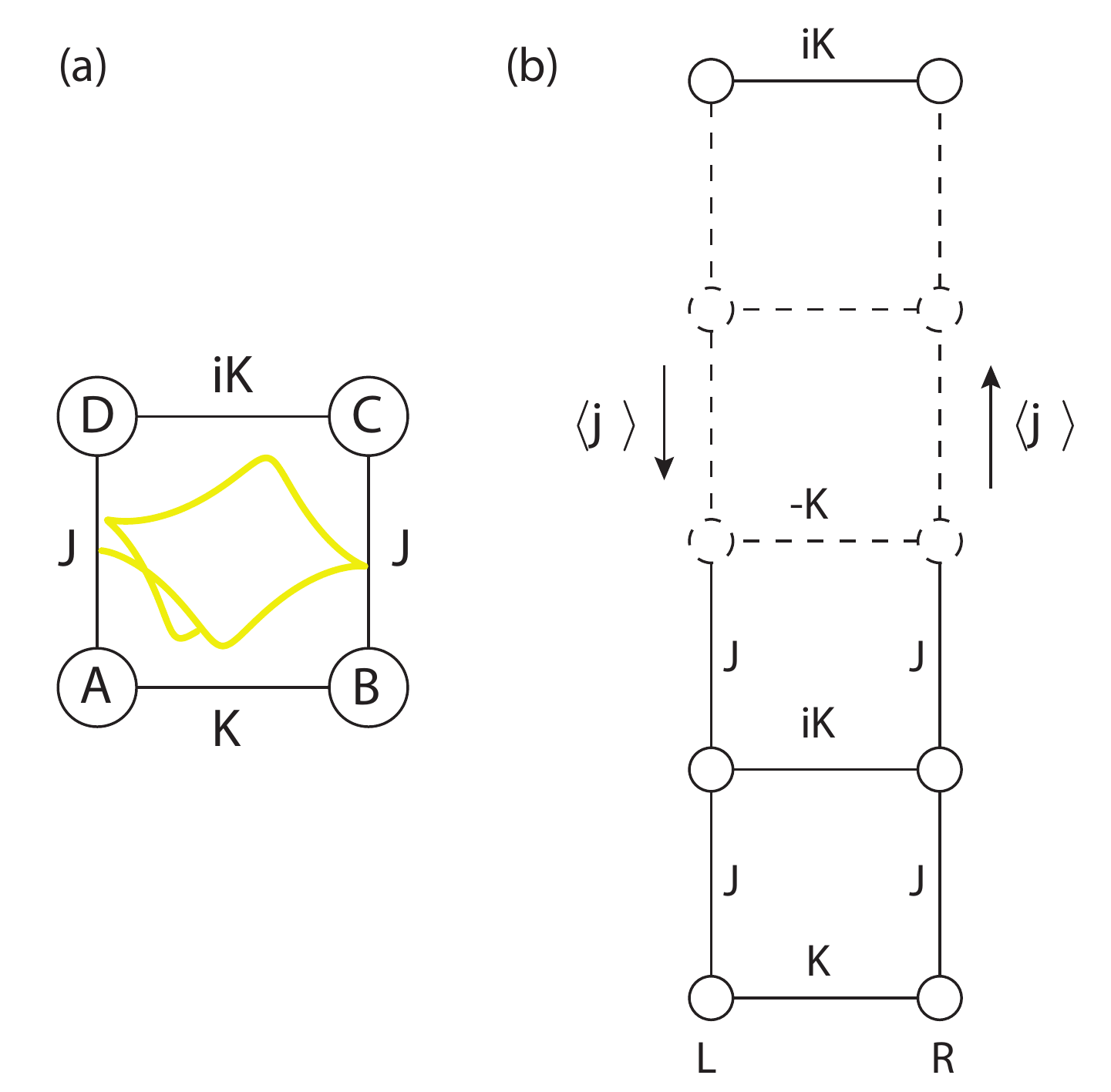}
\caption{\label{fig:plaq_lad}(Color online). Schematic representation of the HH plaquette (panel (a)) with the cyclotron orbit, and the HH ladder (panel (b)) with the chiral currents. }
\end{figure} 

In the end of this section, we would like to set up the stage for the two geometries we shall consider in the following two sections. The first geometry is that of a $2\times 2$ plaquette, c.f.~Fig.~\ref{fig:plaq_lad}, panel (a). The four sites are labelled by the letters $A$ through $D$. In the HH model, a particle prepared in the superposition state $|\psi_0\rangle = 1/\sqrt{2}\left(|A\rangle + |D\rangle\right)$ will follow a quantum cyclotron orbit on average~\cite{aidelsburger_13}. If we label the on-site number operator by $n_j$ ($j = A\dots D$), one can define the average position of the particle along the $x$- and $y$-directions as $\langle X\rangle =  (N_\text{right} - N_\text{left})/2$, and $\langle Y\rangle = (N_\text{up} - N_\text{down})/2$, where $N_\text{left} = n_A + n_D$, $N_\text{right} = n_B + n_C$, $N_\text{up} = n_C + n_D$, $N_\text{down} = n_A + n_B$. The lattice constant is set to unity. Motivated by recent experimental results~\cite{aidelsburger_13}, we are interested in the evolution of the quantum cyclotron orbit $(\langle X\rangle (t),\langle Y\rangle (t) )$ of a single plaquette for finite driving frequencies.

The second geometry we consider is a two-legged ladder~\cite{huegel_14}. In order to keep the discussion consistent with the recent experiment of Ref.~\onlinecite{atala_14}, we position the ladder along the $y$-direction. There is no tilt along this direction, so the hopping elements acquire Peierls phases only in between the two legs, labelled by $m = L,R$. A schematic representation is shown in Fig.~\ref{fig:plaq_lad}, panel (b). For the numerical simulations, we choose a $2\times 20$ ladder with open boundary conditions. We are interested in the evolution of the local current operator along the two legs, denoted by $j_m^{n,n+1}$, where $m$ denotes the left or right leg, and $n$ labels the sites in the vertical (ladder) direction. To avoid finite-size effects, we constrain our discussion to the current flowing between the vertical sites $n=10$ and $n=11$. We work at unit filling.
 
From now on, we focus on the non-interacting model exclusively. 

\section{\label{sec:FS} Floquet Stroboscopic Evolution}

\begin{figure}[b]
\includegraphics[width=\columnwidth]{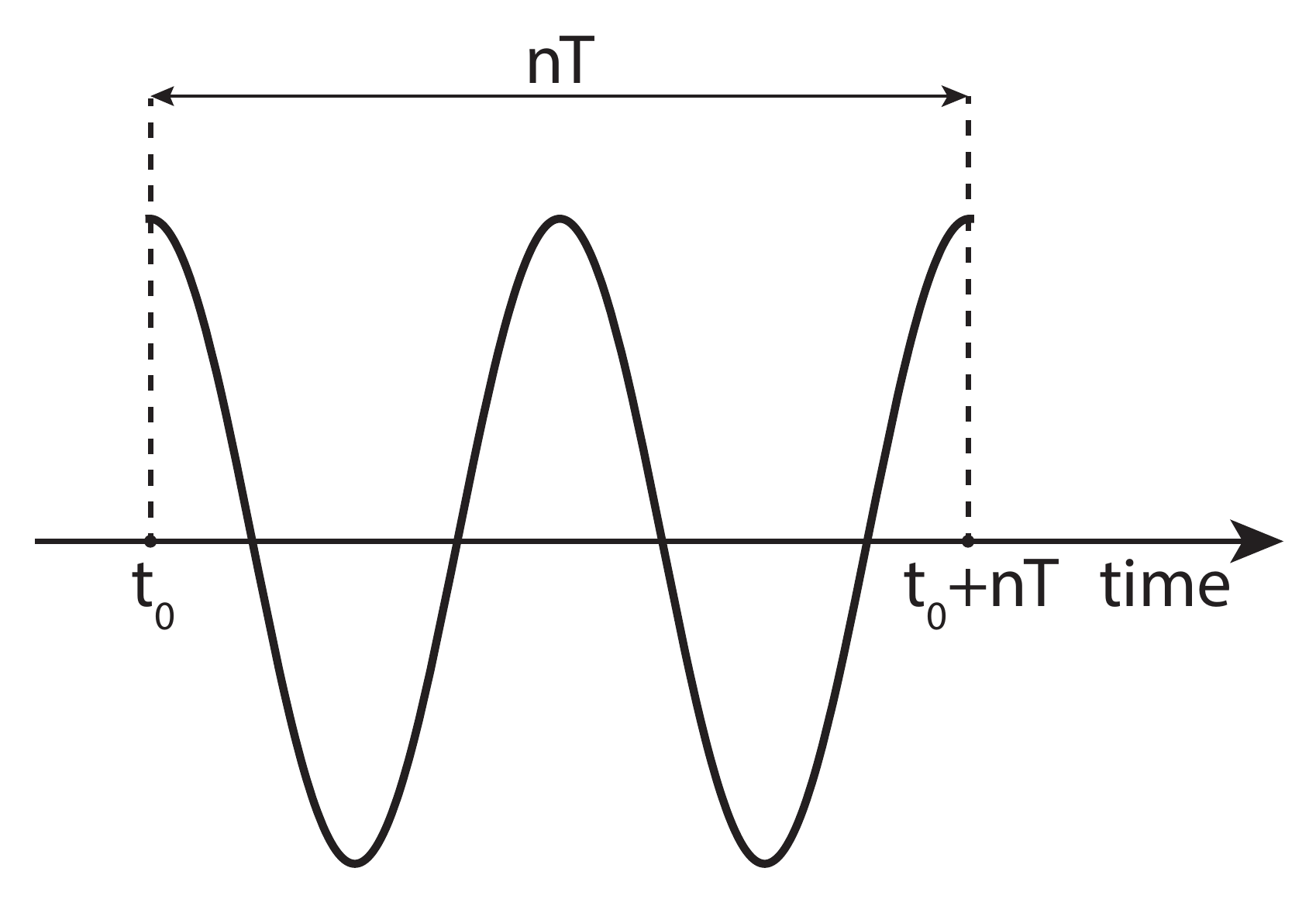}
\caption{\label{fig:FS_meas} Floquet stroboscopic (FS) evolution. Figure taken from Ref.~\onlinecite{bukov_14}. }
\end{figure} 

Let us initialise the periodic driving at time $t=0$. After precisely $n$ driving periods, we stop the evolution at time $nT$, c.f.~Fig.~\ref{fig:FS_meas}. The Floquet Stroboscopic (FS) evolution makes use of the kick-operator identity $P(nT) = \mathds{1}$, c.f.~Eq.~\eqref{eq:floquet_thm}, to define a stroboscopic evolution operator $U(nT,0) = \exp(-i H_F nT)$ w.r.t.~the time-independent Floquet Hamiltonian, $H_F$. Here $n\in \mathbb{N}$ is a positive integer, and $T = 2\pi/\Omega$ is the driving period. The FS evolution is particularly appealing, since it allows one to study a time-independent problem, provided that $H_F$ can be computed in some suitable limit of interest. 

It follows that the FS-evolution of any observable $\mathcal{O}$ is described by 
\begin{eqnarray}
\langle\mathcal{O}\rangle(nT) &=& \langle\psi_0^\text{lab}|e^{iH_F^\text{lab}nT}\mathcal{O}^\text{lab}e^{-i H_F^\text{lab}nT}|\psi^\text{lab}_0\rangle\nonumber\\
&=& \langle\psi_0^\text{rot}|e^{iH_F^\text{rot}nT}\mathcal{O}^\text{rot}(t)e^{-i H_F^\text{rot}nT}|\psi^\text{rot}_0\rangle,
\label{eq:strobo_evolution_gen}
\end{eqnarray}
where $|\psi^\text{rot}_0\rangle  = V(0)|\psi^\text{lab}_0\rangle$ are the initial states in the lab and the rotating frame, respectively. Similarly, $\mathcal{O}^\text{rot}(t) = V(t)\mathcal{O}^\text{lab}V^\dagger(t)$. The two Floquet operators are related by $H_F^\text{rot} = V(0)H_F^\text{lab}V^\dagger(0)$. Here $V(t)$ denotes the transformation into the rotating frame given in Eq.~\eqref{eq:V_rot_frame}.

Notice that the expectation value itself does not depend on whether the analysis is performed in the lab or the rotating frame. Working in the rot frame offers certain advantages w.r.t.~organising a perturbative series expansion~\cite{bukov_14}. In this case, one needs to also transform the initial state, accordingly.

\begin{figure*}[!ht]
\begin{minipage}{\textwidth}
	\includegraphics[width = \textwidth]{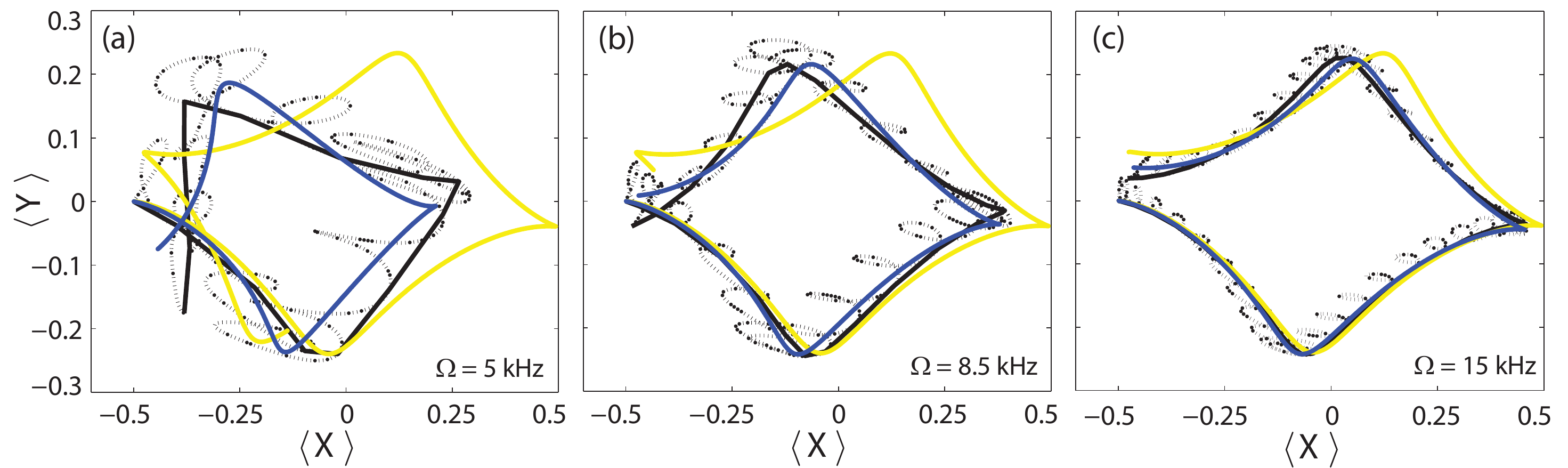}
	\caption{\label{fig:FS_orbits}(Color online). FS vs.~exact dynamics of the quantum cyclotron orbits on a $2\times 2$ plaquette for different values of the driving frequency. The yellow curve is the one expected from the HH model; the grey dotted curve follows the exact evolution w.r.t.~$H^\text{lab}(t)$, while the black curve follows the exact FS evolution. The green curve describes the system evolving w.r.t.~the HH Hamiltonian plus the leading $\Omega^{-1}$-correction. The model parameters are ($\hbar = 1$) $J_x = 0.85$kHz, $J_y = 0.5$kHz, $U=0$kHz, and $\Phi_\square = -\pi/2$, which yields $K = -0.23$kHz, $J = 0.46$kHz, and $\zeta = - 0.57$. The lattice constant is set to unity. }
\end{minipage}
\begin{minipage}{\textwidth}
	\includegraphics[width = \textwidth]{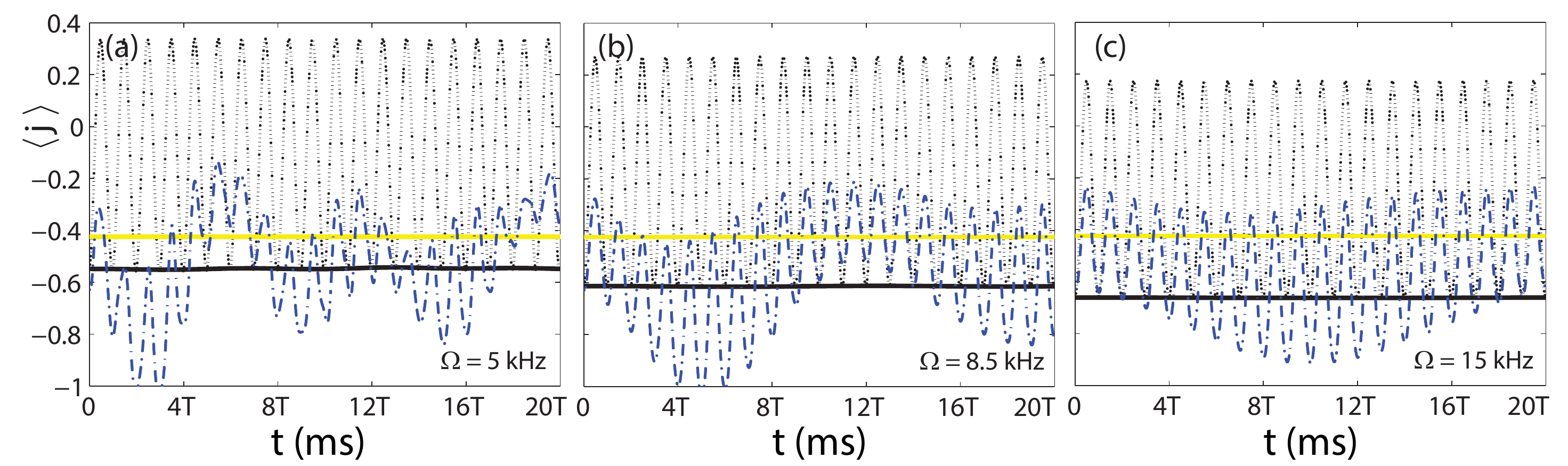}
	\caption{\label{fig:FS_currents}(Color online). FS vs.~exact dynamics of the probability (particle) current expectation $\langle j_L^{10,11} \rangle$ on a $2\times 20$ ladder at unit filling: the yellow curve denotes the evolution w.r.t.~the HH Hamiltonian starting from the ground state (GS) of the HH model. The grey dotted curve is the exact evolution of the lab-frame current evolved with $H^\text{lab}(t)$ starting from the GS of the exact Floquet Hamiltonian, while its stroboscopics at times $nT$ is shown by the black line. The blue dashed-dotted line is the exact evolution of the lab-frame current evolved with $H^\text{lab}(t)$ starting from the GS of the HH Hamiltonian. The model parameters are ($\hbar = 1$) $J_x = 1.2$kHz, $J_y = 0.16$kHz, $U=0$kHz, and $\Phi_\square = -\pi/2$, which yields $K = -0.3$kHz, $J = 0.15$kHz, and $\zeta = - 0.53$; the system is in the Meissner phase~\cite{atala_14}.    }
\end{minipage}	
\end{figure*}

\emph{Quantum Cyclotron Orbits.} Let us first consider the FS evolution of the quantum cyclotron orbits. We prepare the system at time $t=0$ in the state $|\psi_0\rangle = 1/\sqrt{2}\left(|A\rangle + |D\rangle\right)$, and track the time evolution of the particle motion along the trajectory $(\langle X\rangle (t),\langle Y\rangle (t) )$. Figure~\ref{fig:FS_orbits}, panels (a)-(c), shows the quantum cyclotron orbits for driving frequencies $\Omega = 5\text{Hz}, 8.5\text{Hz},$ and $15\text{Hz}$. Let us focus on panel (a). The yellow curve shows the orbit, as expected from the Harper-Hofstadter (HH) Hamiltonian. It corresponds to infinite-frequency driving and, if one aims at simulating the physics of the HH model, a good agreement between the measurement and this curve is expected. The grey dotted curve, on the other hand, shows the exact time evolution w.r.t.~$H^\text{lab}(t)$. For $\Omega=15\text{Hz}$, panel (c), on average it follows closely the $\infty$-frequency curve, with the agreement becoming worse at small frequencies. Note the additional structure - the wiggles - which appear within the driving periods. They become more pronounced at small frequencies, and are a manifestation of the fast evolution due to the kick operator $P(t)$. 

Notice also that the size of the area enclosed by the orbits shrinks as the driving frequency decreases. This phenomenon is due to the fact that, for finite $\Omega$, the driving is no longer resonant with the energies of the tilted double well. Had we chosen the driving frequency to be resonant, i.e.~$\Omega = \sqrt{\Delta^2 + 4J_x^2}$, where $\Delta$ defines the slope of the tilt, on average the orbits would span the same area at any frequency. In general, the shrinking of the orbits is due to diagonal (in the Fock basis) terms in the Hamiltonian. In the case of the present discussion, such terms are generated by the site-dependent chemical potential which enters the Hamiltonian in the form of a $\Omega^{-1}$-correction. If interactions of the same magnitude as the effective hopping are included, they also lead to reduction in the size of the cyclotron orbits~\cite{li_13}. In the case of a resonant drive, this correction is precisely compensated for by a $\Omega^{-1}$-leftover of the tilt in the $x$-direction~\footnote{In order to apply Floquet theory in the case of a resonant drive, one first writes the gradient term as $\Delta = \Omega + \Delta - \Omega\approx \Omega -2J_x^2/\Delta$. Then one goes to the rotating frame w.r.t.~$\Omega$. Whenever $\Omega\sim\Delta$ the second term amounts to a leftover tilt, present also in the rotating frame.}. For a rational flux per plaquette the site-dependent chemical potential is periodic, with period set by the magnetic unit cell. Since the $\Omega^{-1}$-leftover in the tilt is not periodic, this cancellation will not work for systems, containing more than two sites along the direction of the tilt. 

The black line in Fig.~\ref{fig:FS_orbits}, panels (a)-(c) is the FS orbit. Its smoothness depends on the scale of the driving period. It follows that it captures well the exact evolution on average. Finally, the green curve is the evolution w.r.t.~the HH Hamiltonian plus the $\Omega^{-1}$-correction processes, discussed in Sec.~\ref{sec:model_corrections}. Comparing it to the FS curve, we deduce that the first-order correction in this noninteracting model is enough to capture the evolution up to scales $J_x/\Omega\sim 0.2$.

\emph{Chiral Currents.}  We now enlarge the system size and consider a $2\times 20$ ladder, as described at the end of Sec.~\ref{sec:model_corrections}. The observable of interest is the chiral current along the legs of the ladder~\cite{huegel_14}. Our goal is to simulate the HH current in the ground state (GS) of the HH model at finite frequencies.

Let us consider Heisenberg's EOM for the number operator. Using Floquet's theorem, Eq.~\eqref{eq:floquet_thm}, the latter can be cast into the form~\cite{bukov_14}
\begin{eqnarray}
i \partial_t n_{mn}(t)&=& \mathrm e^{i H_F t} [P^\dagger(t) n_{mn} P(t), H_F] \mathrm e^{-i H_F t}\nonumber\\
&+& i \mathrm e^{i H_F t}  \partial_t \left(P^\dagger (t) n_{mn} P(t) \right)\mathrm e^{- i H_F t}.
\end{eqnarray}
The RHS has two contributions: the first term describes the evolution w.r.t.~the Floquet Hamiltonian $H_F$, while the second term contains information about the intra-period evolution governed by the Kick operator $P(t)$. In the high-frequency limit $H_F$ governs the slow, and $P(t)$ - the fast evolution. Taking into account the time-scale separation, notice that, only if one averages the above equation over one driving period, does the second term on the RHS vanish, owing to the periodicity of the Kick operators, and one can define a time evolution solely w.r.t.~$H_F$. By the associated continuity equation~\cite{bukov_14}, there exists a current, conserved under the evolution w.r.t.~$H_F$. On the other hand, the FS evolution discards every information about the fast evolution stored in $P(t)$. Hence, the current associated with the FS evolution cannot possibly be conserved under $H_F$. Consequently, even in the $\Omega\to\infty$ limit where $H_F$ coincides with the HH Hamiltonian, we expect that the FS evolution fails to reproduce the physics of the HH Hamiltonian correctly. Physically, this is related to the fact that, unlike the number operator, the current operator does not commute with the transformation to the rotating frame, $V(t)$, and, therefore, it exhibits strong oscillations within a driving period.

It is important to keep in mind that any experiment is performed in the lab frame, and hence measures the lab-frame current. On the other hand, the HH current is aware of the magnetic field present. The current operators are given by  

\begin{eqnarray}
j_{m}^{n,n+1;\text{lab}} &=& -i J_y a^\dagger_{m,n+1} a_{mn} + \text{h.c.},\nonumber\\
j_{m}^{n,n+1;\text{rot}}(t) &=& -i J_y e^{-i\zeta\sin(\Omega t+ \phi_{mn})} a^\dagger_{m,n+1} a_{mn} + \text{h.c.},\nonumber\\
j_{m}^{n,n+1;\text{HH}} &=& -i J a^\dagger_{m,n+1} a_{mn} + \text{h.c.}.
\label{eq:currents_lab_vs_harper}
\end{eqnarray}
One might be tempted to think that the only difference between the lab-current and the HH-current is the renormalisation of the hopping matrix element. However, this is an illusion which arises due to the gauge choice in the HH model. Had we considered the HH-current along the $x$-direction, one would need to make the replacement $J_x\rightarrow K e^{i\phi_{mn}}$, including the Peierls phase. Hence the lab-frame and the HH-current operators are fundamentally two different objects. 

To take into account finite-frequency effects, we consider two initial states: the GS of the HH Hamiltonian, and the GS of the Floquet Hamiltonian (the former being the $\infty$-frequency limit of the latter). To avoid finite-size effects, we focus on the current flowing from site $n=10$ to $n=11$ on the left leg $m=L$. We work at unit filling.

Figure~\ref{fig:FS_currents}, panels (a)-(c), shows the time-evolution of the current expectation value for a set of three different driving frequencies. The yellow curve is the HH-current expectation value, starting from the GS of the HH Hamiltonian. Since the time-evolution is w.r.t.~the HH Hamiltonian, this curve is constant in time. Moreover, it is also independent of the driving frequency, for the HH model arises in the infinite-frequency limit. Any experiment, which simulates the HH model successfully and is able to measure the chiral currents directly, should reproduce this curve.

The grey dotted curve shows the lab-frame current expectation value, evolved with the exact Hamiltonian $H^\text{lab}(t)$. The initial state is the GS of the Floquet Hamiltonian, which takes into account finite-frequency corrections to the HH Hamiltonian to all orders. Although being due to the kick operator $P(t)$, the secondary oscillations at the short time scale do not vanish in the high-frequency limit in this case. This is because the current operator is not invariant under the transformation in the rotating frame, Eq.~\eqref{eq:V_rot_frame} and, hence, may exhibit large period-to-period fluctuations.

The black curve is the corresponding FS evolution, i.e.~the grey-dotted curve evaluated at times $nT$. Notice how increasing the driving frequency leads to a larger deviation from the yellow line. Hence, it follows that the FS evolution is not suitable for describing the HH-current at any frequency. Theoretically, this follows from the fact that the lab-frame current does not include the magnetic field, generated in the high-frequency limit via the Peierls phase, c.f.~Eq.\eqref{eq:currents_lab_vs_harper}. In fact, the relation between the lab- and the HH-current is similar to the one between canonical and mechanical momentum, which are two distinct operators. Consequently, stroboscopic evolution is incapable of capturing the features of the HH-current. It merely evolves the lab-frame current operator with the Floquet Hamiltonian. We remark that this fact does not question the experimental results of Ref.~\onlinecite{atala_14}, since the HH-current there was not measured in a stroboscopic fashion.

One might be tempted to think that by choosing a different phase of the driving protocol (i.e.~a different Floquet gauge) one might be able to achieve a stroboscopic evolution, such that the black curve follows the average over one period of the grey dotted curve. However, in general, this is not the case, since choosing a different Floquet gauge would not result in the HH Hamiltonian, but in a unitarily equivalent one, and the associated current operator will differ from the HH-current.

Finally, the blue dash-dotted curve in Fig.~\ref{fig:FS_currents}, panels (a)-(c), shows the evolution of the lab-frame current w.r.t.~$H^\text{lab}(t)$ (so far same as grey-dotted curve), but starting from the GS of the HH Hamiltonian. We see that the time-evolution is more complicated in this case. We shall discuss the origin of this more complicated behaviour in the next section.

\section{\label{sec:FNS} Floquet Non-stroboscopic Evolution}

\begin{figure}[b]
\includegraphics[width=\columnwidth]{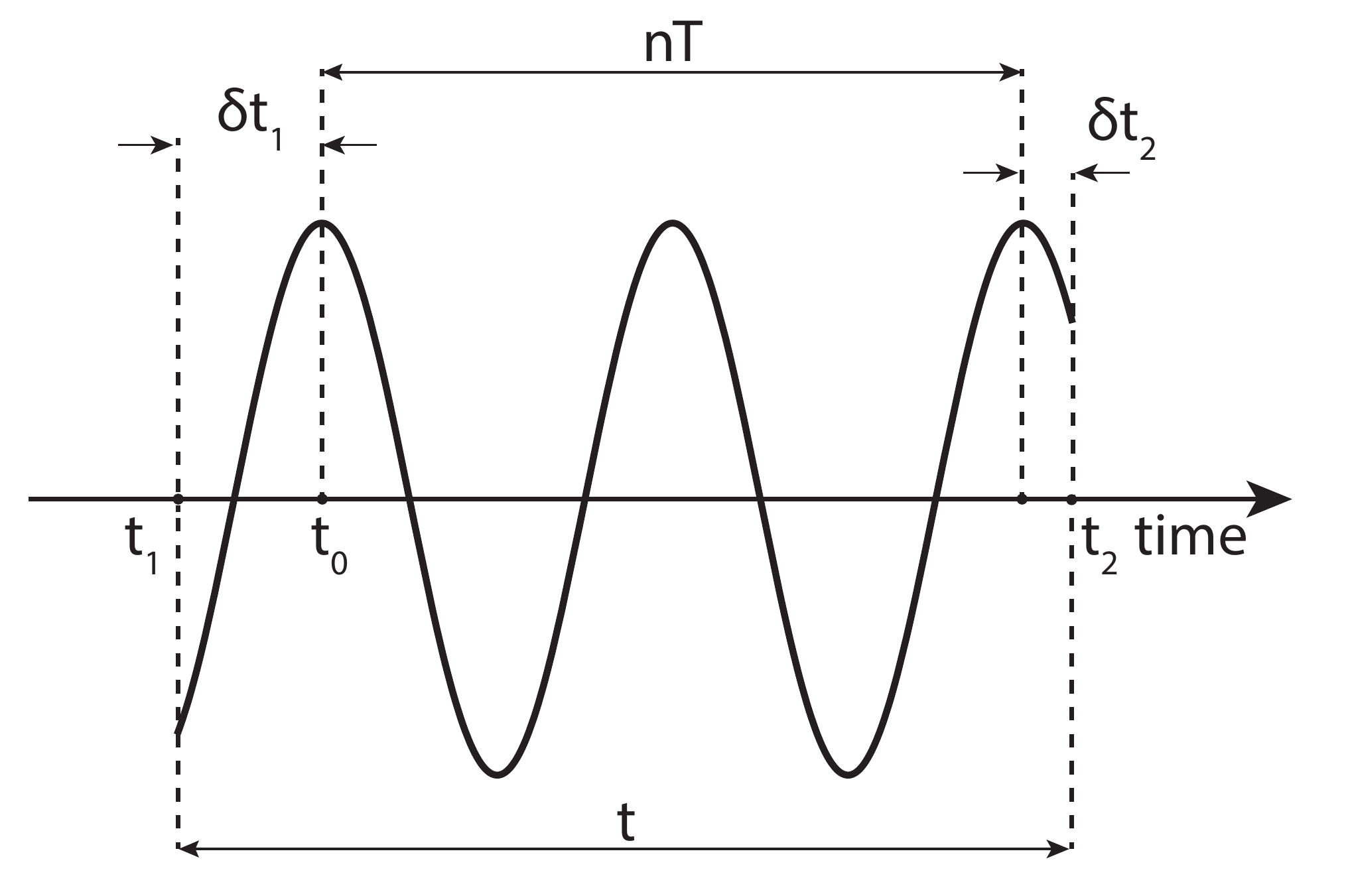}
\caption{\label{fig:FNS_meas} Floquet non-stroboscopic (FNS) evolution. Figure taken from Ref.~\onlinecite{bukov_14}. }
\end{figure}

Let us now consider the Floquet non-stroboscopic (FNS) evolution. We begin by dividing the total time evolution interval $[t_1,t_2]$ in three parts, c.f.~Fig.~\ref{fig:FNS_meas}. First, the driving is switched on at time $t_1$. Second, at time $t_0$, we set up the timer for the stroboscopic frame of $n$ periods, which lasts up to time $t_0 + nT$. Finally, we stop the evolution at time $t_2$ in the $(n+1)^\text{st}$ driving period.  

The concept of the Floquet non-stroboscopic evolution is designed to take into account experimental uncertainties in the phase of the initial driving and the time at which the evolution stops. Suppose that, due to experimental constraints, one cannot control the phase of the driving. If this is the case, the short time interval $\delta t_1 = t_0 - t_1$ between the beginning of the evolution, $t_1$, and the beginning of the stroboscopic frame, $t_0$, changes every time we switch on the driving. Similarly if, in the high-frequency limit, when the driving period is a small number, one cannot resolve the time evolution within a single driving period well, it is appropriate to assume that the short piece of evolution within the $(n+1)^\text{st}$ driving period $\delta t_2$ (see Fig.~\ref{fig:FNS_meas}) varies from one realisation of the experiment to another. 

\begin{figure*}[!ht]
	
\begin{minipage}{\textwidth}
	\includegraphics[width = \textwidth]{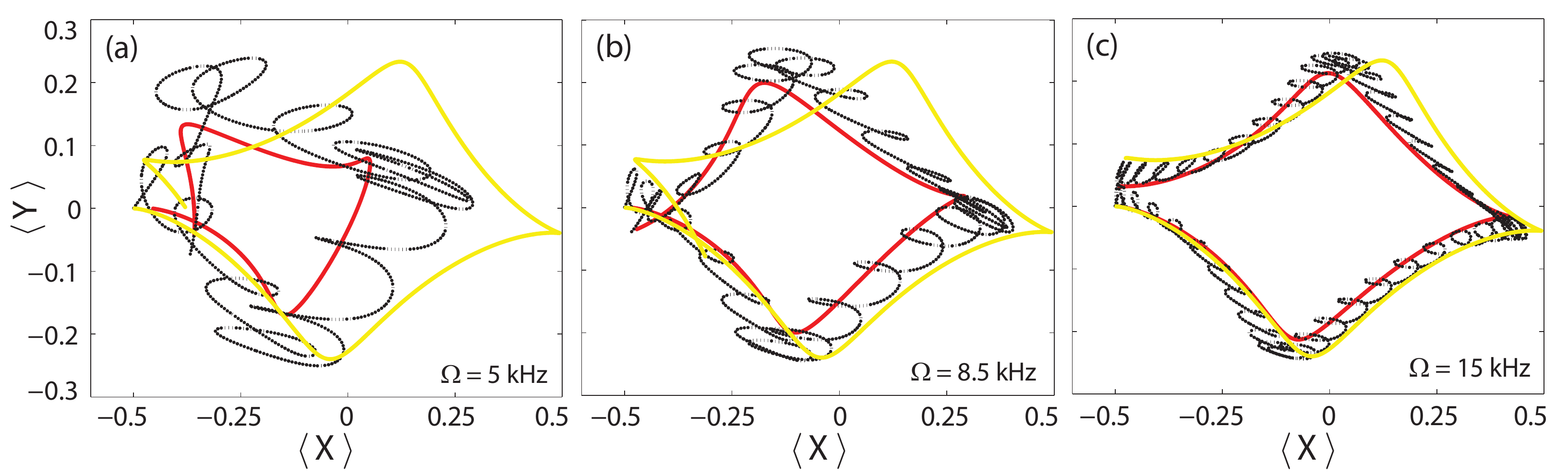}
	\caption{\label{fig:FNS_orbits}(Color online). FNS vs.~exact dynamics of the quantum cyclotron orbits on a $2\times 2$ plaquette for different values of the driving frequency. The yellow and grey dotted curves are the same as in Fig.~\ref{fig:FS_orbits}. The red curve shows the FNS evolution of the cyclotron orbit. The model parameters are ($\hbar = 1$) $J_x = 0.85$kHz, $J_y = 0.5$kHz, $U=0$kHz, and $\Phi_\square = -\pi/2$, which yields $K = -0.23$kHz, $J = 0.46$kHz, and $\zeta = - 0.57$. The lattice constant is set to unity. }
\end{minipage}

\begin{minipage}{\textwidth}
	\includegraphics[width = \textwidth]{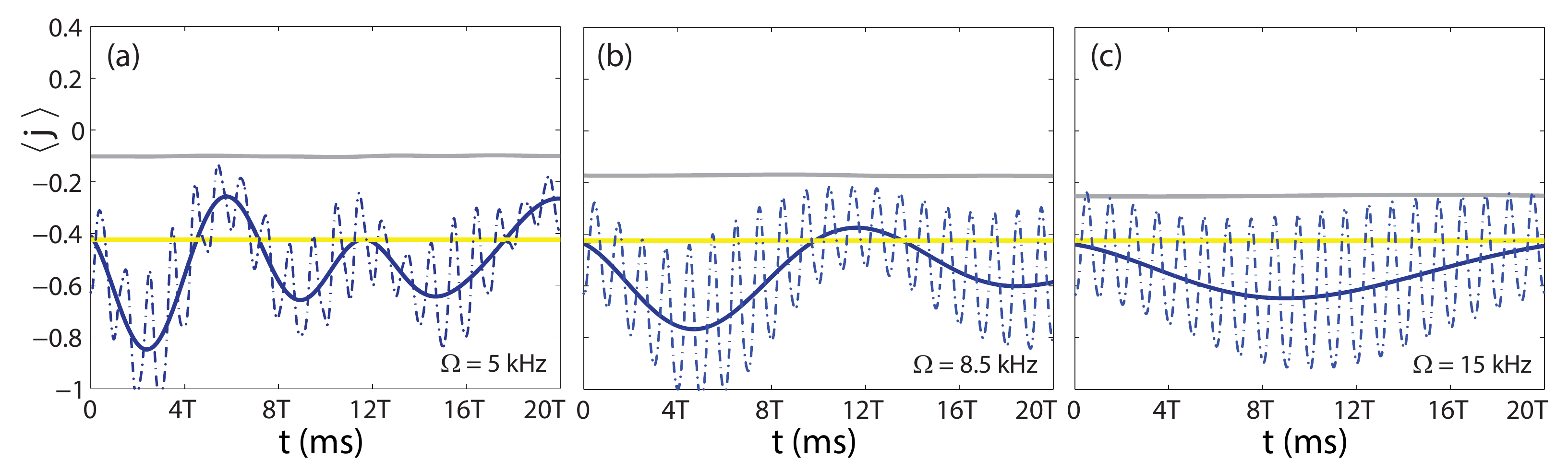}
	\caption{\label{fig:FNS_currents}(Color online). FNS$_2$ vs.~exact dynamics of the probability (particle) current expectation $\langle j_L^{10,11} \rangle$ on a $2\times 20$ ladder at unit filling: the yellow and blue dashed-dotted curves are the same as in Fig.~\ref{fig:FS_currents}. The grey line is the FNS$_2$ evolution of the lab-frame current starting from the GS of the Floquet Hamiltonian, while the solid blue line is the FNS$_2$ evolution of the lab-frame current starting from the GS of the HH Hamiltonian. The FNS$_2$ evolution is always w.r.t.~the Floquet Hamiltonian. The model parameters are ($\hbar = 1$) $J_x = 1.2$kHz, $J_y = 0.16$kHz, $U=0$kHz, and $\Phi_\square = -\pi/2$, which yields $K = -0.3$kHz, $J = 0.15$kHz, and $\zeta = - 0.53$; the system is in the Meissner phase~\cite{atala_14}. }
\end{minipage}
\end{figure*}

One can conveniently take into account this uncertainty in the length of the initial and final driving periods by assuming that $\delta t_{1,2}$ are independent random variables, uniformly distributed over one driving period $T$. One can then average the exact expectation value over $\delta t_1$ and $\delta t_2$. The resulting FNS evolution is equivalent to a statistical expectation value starting from a mixed initial state. The corresponding density matrix $\overline{\rho}$ arises due to averaging over the phase of the driving, or equivalently $\delta t_1$. The observable in the FNS evolution becomes dressed, $\mathcal{O} \to\overline{\mathcal{O}}$, due to averaging over the final time $\delta t_2$. 

\begin{eqnarray}
\langle\mathcal{O}\rangle_{\text{FNS}}(t) &=& \text{tr}\left( \overline{\rho}e^{i H_F t}\overline{\mathcal{O}}e^{-i H_F t}\right),\nonumber\\
\overline{\mathcal{O}} &=& \frac{1}{T}\int_0^T\mathrm{d}\delta t_2 P^\dagger(\delta t_2)\mathcal{O}P(\delta t_2), \nonumber\\
\overline{\rho} &=& \frac{1}{T}\int_0^T\mathrm{d}\delta t_1 P^\dagger(\delta t_1)|\psi_0\rangle\langle\psi_0|P(\delta t_1).
\label{eq:Floquet_meas_general}
\end{eqnarray}
As in the FS evolution, the FNS evolution is solely w.r.t.~the time-independent Floquet Hamiltonian $H_F$. Effects due to the kick operator $P(t)$ are now taken into account by the time-average. The FNS evolution is explained in great detail in Ref.~\onlinecite{bukov_14}. As with normal observables, the dressed quantities can be expressed in the rotating frame via $\overline{\mathcal{O}}^\text{rot} = V(0) \overline{\mathcal{O}}^\text{lab} V^\dagger(0)$, $\overline{\rho}^\text{rot} = V(0) \overline{\rho}^\text{lab} V^\dagger(0)$.

In the infinite-frequency limit, one can establish the following general rule of thumb for the dressed density matrix and observables. If an operator commutes with the driving Hamiltonian $H_\text{drive}(t)$ at all times, then in the limit $\Omega\to\infty$ the corresponding dressed operator is not modified w.r.t~the undressed one. This is intimately related to the fact that the transformation in the rotating frame $V(t)$ also commutes with it. Therefore, the dressed local number operator does not get any modifications in the $\infty$-frequency limit. On the other hand, the lab-frame current operator does not commute with $H_\text{drive}(t)$. As a consequence, it will become dressed and, in the infinite-frequency limit, the dressed lab-frame current operator coincides precisely with the HH-current. Whether the density matrix becomes dressed or not, is determined by whether the initial state is an eigenstate of the operator $H_\text{drive}(t)$.    

The dressed density matrix and observables can be expanded in the high-frequency limit in a perturbative series in powers of the inverse frequency~\cite{bukov_14}. In general, corrections lead to delocalisation. For the density matrix, this means that finite-frequency effects lead to non-zero matrix elements which further contribute to the mixed character of the effective initial state. The corrections to the dressed observables also lead to delocalisation: for instance, the dressed local number operator $\overline n_{mn}$ to order $\Omega^{-1}$ will contain operators similar to the local current between the site $(m,n)$ and all its adjacent neighbours, but these current-like corrections come with renormalised coefficients. The situation is similar for the corrections of the dressed current operator. For convenience, we refrain from showing such lengthy expressions in this work.

\emph{Quantum Cyclotron Orbits.} Let us now revisit the plaquette geometry and consider the FNS evolution of the quantum cyclotron orbits. Since the initial state $|\psi_0\rangle = 1/\sqrt{2}\left(|A\rangle + |D\rangle\right)$ is not an eigenstate of $H_\text{drive}(t)$, the density matrix $\overline \rho$ in the $\infty$-frequency limit gets dressed.

\begin{align}
\overline{\rho_0^\text{rot}} \stackrel{\Omega\to\infty}{=}& \frac{1}{2}\bigg(|A\rangle\langle A| + |D\rangle\langle D| +  \mathcal{J}_0(\zeta) |A\rangle\langle D| + \text{h.c.}  \bigg),\nonumber\\
\overline{n^\text{rot}}_{mn} \stackrel{\Omega\to\infty}{=}& n_{mn}.
\end{align}
It follows that $\overline{\rho}$ represents a mixed state at any $\zeta$, such that $\mathcal{J}_0(\zeta)\neq 0$. In this sense, the FNS expectation value is somewhat reminiscent of finite-temperature, although we could not establish a general relation between the two.

Figure~\ref{fig:FNS_orbits}, panels (a)-(c), compares the FNS evolution (red) to the exact evolution (grey dotted line), and the evolution expected from the HH model (yellow line). For the FNS evolution curve, we used the exact dressed observable and density matrix, calculated to all orders in the inverse frequency, and the evolution is performed w.r.t.~the Floquet Hamiltonian $H_F$. Due to the change of the initial condition from a pure state to a mixed state, the orbit no longer starts from the expected point $(\langle X\rangle, \langle Y \rangle) = (-0.5, 0)$. The shrinking in the size of the orbit is due to the non-resonant drive, c.f.~Sec.~\ref{sec:FS}. This, together with the delocalisation of the local number operator at finite frequencies, leads to a further deviation of the orbit from the time-averaged one. 

Nevertheless, we can conclude that when measuring the local density and combinations thereof, the FNS evolution is equally good as the FS evolution, especially in the high-frequency limit. We expect this to hold for any observable which commutes with $H_\text{drive}(t)$.

\emph{Chiral Currents.} On the other hand, the situation is very different for the FNS evolution of the chiral currents. It follows that, in this case, the infinite-frequency FNS-current coincides precisely with the HH-current:

\begin{equation}
\overline{j}_{m}^{n,n+1;\text{rot}}  \stackrel{\Omega\to\infty}{=} \frac{1}{T}\int_0^T\mathrm{d}\delta t_2 j_{m}^{n,n+1;\text{rot}}(\delta t_2) = j_{m}^{n,n+1;\text{HH}}.
\end{equation} 
Hence, we expect that the FNS expectation of the lab-frame current correctly reproduces the $\infty$-frequency behaviour of the HH-current. 

In order to explore this in greater detail, let us assume that the phase of the driving can be fixed, so that there is no need to average over the initial time $\delta t_1$. Hence, we need not dress the density matrix, and the initial state remains a pure state. At the same time, we assume that we do not have a perfect control over the final time $\delta t_2$. Therefore, we still have to dress the current operator. We refer to this type of non-stroboscopic evolution as FNS$_2$.   

Figure~\ref{fig:FNS_currents}, panels (a)-(c), compare the FNS$_2$ evolution of the current operator at different driving frequencies. The yellow curve shows the evolution of the HH-current w.r.t.~the HH Hamiltonian, in the GS of the HH Hamiltonian. This is the same curve as in Fig.~\ref{fig:FS_currents}, panels (a)-(c). An ideal simulation of the HH model would reproduce this curve. 

The grey curve is the evolution of $\overline{j}_{m}^{n,n+1;\text{rot}}$, w.r.t.~HH Hamiltonian and in the GS of the HH Hamiltonian. The FNS-current here is calculated numerically including all inverse-frequency corrections. Observe that, contrary to the FS evolution, this curve (grey) approaches the curve expected from the HH model (yellow) in the limit $\Omega\to\infty$.

The blue curve is the same as the grey one, but starting from the GS of the exact Floquet Hamiltonian, $H_F$. For comparison, we also show the blue dash-dotted curve from Fig.~\ref{fig:FS_currents} which shows the corresponding exact evolution. Obviously, it follows that the FNS dynamics captures all the main characteristics of the slow evolution w.r.t.~$H_F$. It simply averages out the fast oscillations due to the kick operator $P(t)$. Hence, in models where the Floquet Hamiltonian has prescribed engineered properties~\cite{bukov_14}, it is the FNS evolution that reveals the physics behind $H_F$, and not the FS one.

\section{\label{sec:conclusion} Conclusion}

We discuss two different types of expectation values in periodically driven systems: the FS and FNS evolution. The FNS expectation value is a statistical one, and the associated density matrix depends mostly on the properties of the initial state (localized vs.~delocalized). We found that, in general, only the FNS evolution gives access to local observables which correspond to the Floquet Hamiltonian. In this context, we showed that stroboscopic evolution will fail to reproduce the dynamics of the Harper-Hofstadter current at any frequency, since it merely evolves the lab-frame current, which is not conserved w.r.t.~the evolution due to $H_F$, with the Floquet Hamiltonian. 

Stroboscopic measurements can reveal the physics of the Floquet Hamiltonian, but only for observables, invariant under the driving Hamiltonian $H_\text{drive}(t)$. For all other types of observables, one needs to resort to FNS evolution. We verified these predictions by examining the quantum cyclotron orbits on a $2\times 2$ plaquette, as well as the evolution of the local current on a $2\times 20$ ladder in the non-interacting system with open boundary conditions. Moreover, in the high-frequency limit, the stroboscopic and the Floquet expectations coincide if, and only if, the initial state is an eigenstate of, \emph{and} the observable commutes with the operator to which the periodic driving couples.

\begin{acknowledgments}
The authors would like to thank M.~Atala and L.~D'Alessio for numerous insightful and 
interesting discussions. This work was supported by NSF DMR-0907039 and AFOSR FA9550-13-1-0039.
\end{acknowledgments}

\end{document}